\pdfoutput=1
\documentclass[preprint,12pt]{elsarticle}
\usepackage{amsmath}
\usepackage{amssymb}
\usepackage{graphicx, color}
\usepackage[nodots,nocompress]{numcompress}

\begin{document}

\begin{frontmatter}

\title{Beam splitter and entanglement created with the squeezed coherent states of the Morse potential}

\author{A Hertz$^1$}
\ead{anaelle.hertz@umontreal.ca}

\author{V Hussin$^2$}
\ead{veronique.hussin@umontreal.ca}

\author{H Eleuch$^{3,4}$}
\ead{heleuch@fulbrightmail.org}

\address{$^1$D\'epartement de Physique,
Universit\'{e} de Montr\'{e}al, Montr\'{e}al, Qu\'ebec, H3C 3J7, Canada}
\address{$^2$D\'epartement de Math\'ematiques et de
Statistique,
Universit\'{e} de Montr\'{e}al, Montr\'{e}al, Qu\'ebec, H3C 3J7, Canada}
\address{$^3$ D\'epartement d'Informatique et de Recherche Op\'erationnelle, Universit\'{e} de Montr\'{e}al, Montr\'{e}al, Qu\'ebec, H3C 3J7, Canada}
\address{$^4$ D\'epartement de G\'enie \'Electrique, \'Ecole Polytechnique de Montr\'eal, Qu\'ebec, H3C 3A7, Canada}


\begin{abstract}

The Morse potential is relatively closed to the harmonic oscillator quantum system. Thus, following the idea used for the latter, we study the possibility of creating entanglement using squeezed coherent states of the Morse potential as an input field of a beam splitter. We measure the entanglement with the linear entropy for two types of such states and we study the dependence with the coherence and squeezing parameters. The new results are linked with observations made on probability densities and uncertainty relations of those states. The dynamical evolution of the linear entropy is also explored.

\end{abstract}

\begin{keyword}
Beam splitter \sep Entanglement \sep Squeezed Coherent states \sep Morse potential


\end{keyword}

\end{frontmatter}


\section{Introduction}

Quantum entanglement is one of the fundamental cornerstones for the quantum information processing. It is this phenomenon that gives to the quantum information all its strength. A lot of research contributions have been done in the last two decades in order to create entangled states that will be used in a future quantum computer. Recently, many devices have been proposed in order to create entangled states \cite{chen}. We can name, for example, cavity QED \cite{Kim1, Xiao}, NMR \cite{Eskandari, Soares}, semiconductor microcavity \cite{el1}, nanoresonator \cite{aty1} or beam splitter \cite{kim02, berrada11}. In this work we will concentrate on this last device.

Squeezed coherent states (SCS) are known since a long time now \cite{walls, Loudon,Dell, gazeau09}. They are very useful in physics due to their property to be quasi-classical, i.e., they minimize the Heisenberg uncertainty relation. SCS of the harmonic oscillator are very well known, and recently, they have been generated for other systems \cite{gazeau09, walls08, Eberle}. In particular, we have recently \cite{angelova08, angelova12} constructed the SCS of the Morse potential, a potential that is a better approximation than the one of the harmonic oscillator for studying the vibrations in a diatomic molecule. This  potential has the particularity to lead to a discrete spectrum of energies which is finite.

SCS of the one-dimensional harmonic oscillator have been used in several experiments \cite{gazeau09, walls08} and, for example, in a beam splitter in order to generate entanglement \cite{kim02, sanders, Our}. 

In this paper, we will follow the same idea, by studying the propagation of the SCS of the Morse potential through a beam splitter. In Section 2, we will resume the construction of  SCS  as eigenstates of a linear combination of ladder operators which are associated to a generalized Heisenberg algebra. Different types of states may be constructed and they corresponds to deformation of the usual SCS. In Section 3, we introduce the known operator $\hat{B}(\theta)$ that defines the beam splitter and a special measurement of entanglement given as the linear entropy $S$. We will also summarize the corresponding results for the case of the harmonic oscillator introducing two types of SCS. Section 4 deals with the case of the Morse potential which has a discrete finite spectrum.  Two different types of SCS, called oscillator-like and energy-like, are under study with respect to the measure of entanglement. In particular, we show that entanglement could be present even if there is no squeezing. The behaviour of the linear entropy explains some of the observations made on probability densities and uncertainty relations of those states in a previous work. The dynamical evolution of the linear entropy is also studied. We conclude the paper in Section 5.

\section{Squeezed coherent states and generalized Heisenberg algebra}

SCS of a one-dimensional quantum system could be constructed as solutions of the eigenvalue equation
\begin{equation}
 (A+\gamma A^{\dag})\Psi(z, \gamma)=z\ \Psi(z, \gamma), \qquad z,\gamma\in \mathbb{C},
 \label{scsdef}
 \end{equation}
where $z$ is the amplitude of coherence  and $\gamma$ the squeezing parameter. The operators $A^{\dag}$ and $A$ are the ladder (creation and annihilation) operators  of the system under consideration and are defined, in general, as:
\begin{equation}
A^{\dag}| n\rangle=\sqrt{k(n+1)}\ | n+1\rangle,\quad  \ A |n\rangle=\sqrt{k(n)}\ | n-1\rangle,
\label{ladder0}
\end{equation}
where  the set  $\{ |n\rangle, n=0,1,...\}$ represents the eigenstates of the energy. We have some freedom in the choice of the positive function $k(n)$ and it will lead to different types of SCS.

If the discrete energy spectrum is infinite (like for the  harmonic oscillator potential), the SCS are exact solutions of (\ref{scsdef}). If the discrete energy spectrum is finite (like for the Morse potential), then the SCS are almost solutions of (\ref{scsdef}). Indeed, the difference between the left and right hand sides of (\ref{scsdef}) is negligible  \cite{angelova12}. In all cases, we can write the SCS as
\begin{equation}
\Psi(z, \gamma)=  \frac{1}{\sqrt{{\cal{N}} (z,\gamma)}}
\sum_{n=0}^{M} \frac{Z(z,\gamma,n)}{\sqrt{\rho(n)}}|n\rangle,
\label{scs}
\end{equation}
where $\rho(n)$ is
\begin{equation}
\rho(n)=\prod_{i=1}^n k(i), \hspace{30pt} \rho(0)=1.
\label{rho}
\end{equation}
The normalization factor is
 \begin{equation}
{\cal N} (z, \gamma)=\sum_{n=0}^{M}{{|Z(z,\gamma,n)|^2}\over{\rho(n)}}.
 \end{equation}

In the expression (\ref{scs}), the limit $M$ of the sum depends on the fact that the energy spectrum is finite or not and $Z(z,\gamma,n)$ satisfies the recurrence relation
\begin{equation}
Z(z,\gamma,n+1)-z Z(z,\gamma,n)+\gamma\  k(n) Z(z,\gamma,n-1)=0 \hspace{30pt} n=1,2,...
\label{recur}
\end{equation}
with $Z(z,\gamma,0)=1$ and $Z(z,\gamma,1)=z$.

In order to make the connection with other approaches  \cite{Hassouni, Abdel}, we will refine the definition of $k(n)$ as
\begin{equation}
k(n)=n\,(f(n))^2.
\label{kn}
\end{equation}
The natural first choice $f(n)=1$ is called oscillator-like type and we get $\rho(n)=n!$ which is essentially the product of the energies of the harmonic oscillator. In this case, we also know that the recurrence relation (\ref{recur}) leads to
\begin{equation}
Z_o(z,\gamma,n)=Z_{HO}(z,\gamma,n)=\Big(\frac{\gamma}{2}\Big)^{\frac{n}{2}} \mathcal{H} \left( n,{\frac{z}{\sqrt{2\gamma}}}\right),
\label{zoh}
\end{equation}
where the subscript $\it{o}$ refers to this type of states and $\mathcal{H} (n,w)$ are the Hermite polynomials. We could also consider the choice $f(n)=\sqrt{v+n}, \ v\geq 0$. This case has been treated for potentials with infinite discrete spectrum like, for example, the trigonometric P\" oschl-Teller potential \cite{PT,Antoine}. We could also consider the choice $f(n)=\sqrt{v-n}, \ v\geq 1$. This case has been treated for potentials like the Morse and  hyperbolic P\" oschl-Teller potentials \cite{Levai} . For special values of $v$, we find $\rho(n)$ as a product of the quadratic energies of the quantum system under consideration that could have infinite or finite spectrum.

Let us mention that all these choices lead to generalized Heisenberg algebras as introduced in the recent literature  \cite{Hassouni, Abdel}. It is usually called a $f$-deformed algebra because of the introduction of the function $f(N)$ in the definition of the ladder operators. Indeed, if we introduce the usual number operator $N$ such that $N| n\rangle=n\ | n\rangle$, we get
\begin{eqnarray}
 [N,A]&=&-A,\qquad [N,A^{\dag}]\,\,=\,\,A^{\dag},\nonumber \\
 \left[ A,A^{\dag}\right]&=&(N+1) (f(N+1))^2-N (f(N))^2.
 \label{gheisen}
\end{eqnarray}
The ladder operators $A^{\dag}$ and $A$ may be related to the ones $a^{\dag}$ and $a$ of the harmonic oscillator by
\begin{eqnarray}
A=af(N)=f(N+1)a,\nonumber\\
A^{\dag}=af(N)a^{\dag}=a^{\dag}f(N+1).
\end{eqnarray}

For $f(n)=1$, the set  $\{A, A^{\dag}, N\}$ closes the usual Heisenberg algebra. For $f(n)=\sqrt{v+\epsilon\ n}$ with $\epsilon =\pm 1$,  we can define new operators  $J_-=A$, $ J_+=A^{\dag}$, $ J_0= N+\frac12 (\delta v+1)$. They generate a $su(1,1)$ algebra for $\delta=1$ or a $su(2)$ algebra for $\delta=-1$, since we have
\begin{eqnarray}
 [J_0,J_-]&=&-J_-,\quad [J_0,J_+]=J_+,\nonumber \\
 \left[ J_-,J_+\right]&=&2 \delta J_0.
 \label{jheisen}
\end{eqnarray}

Thus, if we take the Morse potential ($\epsilon=-1$), comparing (\ref{gheisen}) with (\ref{jheisen}),we get $\delta =-1$, i.e., a $su(2)$ algebra. However, for an infinite potential ($\epsilon=1$), we get  $\delta=1$, i.e., a $su(1,1)$ algebra.

\section{Beam splitter and a measure of entanglement}

\subsection{Beam splitting transformation }

A beam splitter is an optical device which can generate quantum entanglement with a bipartite input state composed of a state $| \psi\rangle$ at one input and a vacuum state $|0\rangle$ at the other input. The beam splitter 50:50 is a necessary part of almost all optical experiments  \cite{walls08} since it allows to split the incident intensity to equal reflected and transmitted intensities. Furthermore, a beam splitter can be used to create entangled states \cite{sanders,Our}. These states have potential applications in quantum cryptography and quantum teleportation \cite{Jennewein,Bennett}.

The effect of a beam splitter can be described by an unitary operator $\hat{B}(\theta)$ connecting the input state with the output one \cite{kim02}:
\begin{equation}
|out\rangle=\hat{B}(\theta) \ |in \rangle= \exp \left[\frac{\theta}{2}(a^{\dag}be^{i\phi}-ab^{\dag}e^{-i\phi})\right] \ |in \rangle,
\end{equation}
where the input state is given as
\begin{equation}
|in \rangle= |\psi\rangle\otimes|0\rangle.
\end{equation}
The ladder operators  $a^{\dag}, a$ and $b^{\dag}, b$ act on the bipartite fields and satisfy $[a, a^{\dag}]= [b, b^{\dag}]=1$. The quantity $\phi$ is the phase difference between the reflected and transmitted fields.

The effect of this operator on the usual Fock state, written as $|n\rangle\otimes|0\rangle$, is given by:
\begin{equation}
\hat{B}(\theta) \big(|n\rangle\otimes |0\rangle\big) =\sum_{q=0}^n \binom{n}{q}^\frac{1}{2} t^q r^{(n-q)}\ |q\rangle\otimes|n-q\rangle.
\label{operateurB}
\end{equation}
The quantities $t$ and $r$ are the transmissibility and reflectivity of the beam splitter that obey the normalization condition $|t|^2+|r|^2=1$. They are related to the angle $\theta$ of the beam splitter by the following equations:
\begin{equation}
t=\cos(\theta/2) \hspace{10pt}\text{and} \hspace{10pt}  r=-e^{-i\phi} \sin(\theta/2).
\label{TandR}
\end{equation}
Later for the calculations and plots, we will use a 50:50 beam splitter, for which we will thus take $\theta=\pi/2$, but the equations in the following are given for the general case.

If the input state is chosen such that $|\psi\rangle$ is a SCS as defined in (\ref{scs}), we get:
\begin{eqnarray}
|out\rangle & = & \hat{B}(\theta)\big( \Psi(z, \gamma)\otimes|0\rangle\big)\nonumber \\
	 &=& \frac{1}{\mathcal{ \sqrt{N} }}\sum_{n=0}^{M} \frac{Z(z,\gamma,n)}{\sqrt{n!}f(n)!}\hat{B}(\theta)\big(|n\rangle \otimes |0\rangle\big)\nonumber \\
	&=&  \frac{1}{\mathcal{ \sqrt{N} }}\sum_{n=0}^{M} \sum_{q=0}^n \frac{Z(z,\gamma,n)}{\sqrt{n!}f(n)!} \binom{n}{q}^\frac{1}{2} t^q r^{(n-q)} |q\rangle\otimes|n-q\rangle\nonumber \\
	&=&  \frac{1}{\mathcal{ \sqrt{N} }}\sum_{q=0}^{M} \sum_{m=0}^{M-q} \frac{Z(z,\gamma,m+q)}{\sqrt{q!m!}f(m+q)!}  t^q r^{m} |q\rangle\otimes|m\rangle.\nonumber \\
	\label{outscs}
\end{eqnarray}

For usual coherent states (CS) of the harmonic oscillator ($Z(z,0, n)=z^n$, $f(n)=1$ and $M=\infty$), it is well-known that the output state (\ref{outscs}) is a tensor product of two CS and so no entanglement is created. However, for SCS ($\gamma\neq 0$), there is always entanglement and we can measure it.  Since different SCS can be constructed for exactly solvable quantum systems, we will  measure the level of entanglement for each type of states.

The beam splitter allowing the propagation of $n$ and $m$ integer numbers of photons in both output channels, we keep the same definition (\ref{operateurB}) of the beam splitter operator $ \hat{B}(\theta)$ acting on the eigenstates of the Morse potential.


\subsection{Linear entropy as a quantification of entanglement}

There exists several measurements of entanglement  such as the concurrence \cite{Wotters}, the negativity \cite{Zyczkowski,Vidal} or the von Neumann entropy \cite{Schumacher,Jozsa,Hausladen,Barnum}. In this article, we chose the linear entropy \cite{Affleck, Gerry}, an upper born of the von Neumann entropy. It is easier to compute, but gives a good indication on the degree of entanglement.

Starting with the density operator $\rho_{ab}$ of a given output state as introduced before, the linear entropy $S$ is defined as
\begin{equation}
S=1-Tr(\rho_a^2),
\end{equation}
where $\rho_a$ is the reduced density operator of the system $a$ obtained by performing a partial trace over system $b$ of the density operator $\rho_{ab}$. The quantity $S$ takes always values between 0 and 1 where 0 corresponds to the case of a pure state and 1 indicates that the considered states have a maximum of entanglement.

For an output state $|out\rangle$  created with a SCS as an input through a beam splitter, the density operator is
\begin{eqnarray}
\rho_{ab} &=& |out\rangle\langle out|\nonumber\\
	&=&\frac{1}{\mathcal{N}}\sum_{q=0}^{M}\sum_{s=0}^{M}\sum_{m=0}^{M-q}\sum_{n=0}^{M-s} \frac{Z(z,\gamma,m+q)}{\sqrt{q!m!}f(m+q)!}\frac{\overline{Z(z,\gamma,n+s)}}{\sqrt{n!s!}f(n+s)!}\nonumber\\
	& & \times \,t^q\, \overline{t^s}\, r^m\, \overline{r^n}\, |q\rangle|m\rangle\langle s|\langle n|
\end{eqnarray}
and the partial trace is
\begin{equation}
\rho_{a} =\frac{1}{\mathcal{N}}\sum_{q=0}^{M}\sum_{s=0}^{M}\sum_{m=0}^{M-q}\frac{Z(z,\gamma,m+q)}{\sqrt{q!}f(m+q)!}\frac{\overline{Z(z,\gamma,m+s)}}{\sqrt{s!}f(m+s)!} \,t^q\, \overline{t^s}\, \frac{|r|^{2m}}{m!}\, |q\rangle\langle s|.
\end{equation}
Thus, the linear entropy becomes
\begin{eqnarray}
S &=& 1- \frac{1}{\mathcal{N}^2}\sum_{q=0}^{M}\sum_{j=0}^{M}\sum_{m=0}^{M-max(q,j)}\sum_{n=0}^{M-max(q,j)}|t|^{2(q+j)} |r|^{2(m+n)}\nonumber\\
 & & \times \, \frac{Z(z,\gamma,m+q)\overline{Z(z,\gamma,m+j)}Z(z,\gamma,n+j)\overline{Z(z,\gamma,n+q)}}{q!j!m!n!f(m+q)!f(m+j)!f(n+j)!f(n+q)!}.
\label{entropy}
\end{eqnarray}

Note that here, the phase $\phi$ between the transmitted and reflected fields is not relevant. Indeed, in the calculation of $S$, the transmissibility $t$ and reflectivity $r$ only appear with the square of their norm.

\subsection{Harmonic oscillator squeezed coherent states and measure of entanglement }

The usual SCS of the harmonic oscillator are defined as (\ref{scs}) where $Z(z,\gamma,n)$ is given by the expression $ (\ref{zoh})$.
The behaviour of theses states is well-known. In our work, we analyze, in particular, the behaviour of the uncertainty relation and the density probability for two types of SCS, i.e., the usual ones for which $f(n)=1$ and the quadratic ones where we take $f(n)=\sqrt{n}$. This last choice is called quadratic since it leads to $k(n)= n^2$ in (\ref{kn}) and it corresponds to an example of states constructed from a $f$-deformed Heisenberg algebra. 

As well-known, the position $x$ and momentum $p$ may be written as a simple combination of the usual ladder operators $a$ and $a^\dag$: 
\begin{equation}
x=\sqrt{\frac{\hbar}{2m\omega}}(a^{\dag}+a) \quad\quad p=i \,\sqrt{\frac{\hbar m\omega}{2}}(a^{\dag}-a).
\end{equation}
Also, the mean value of an arbitrary observable $\mathcal{O}$ in the SCS is $\langle\mathcal{O}\rangle(z,\gamma;t)=\langle\Psi(z,\gamma,x;t)|\mathcal{O}|\Psi(z,\gamma,x;t)\rangle$ where the time evolution of the SCS is given from
\begin{equation}
\Psi(z, \gamma, x;t)=  \frac{1}{\sqrt{{\cal{N}} (z,\gamma)}}
\sum_{n=0}^{\infty} \frac{Z(z,\gamma,n)}{\sqrt{\rho(n)}} e^{-i\frac{E_n}{\hbar}t} \psi_n (x),
\label{scstime}
\end{equation}
where the $\psi_n (x)=\langle x|n\rangle$ are the usual Hermite function in $x$. From this, we get the definition of the dispersion of this observable : $(\Delta \mathcal{O})^2= \langle\mathcal{O}^2\rangle-\langle\mathcal{O}\rangle^2$.

\begin{figure}[!h]
\centering
\includegraphics[width=.7\textwidth]{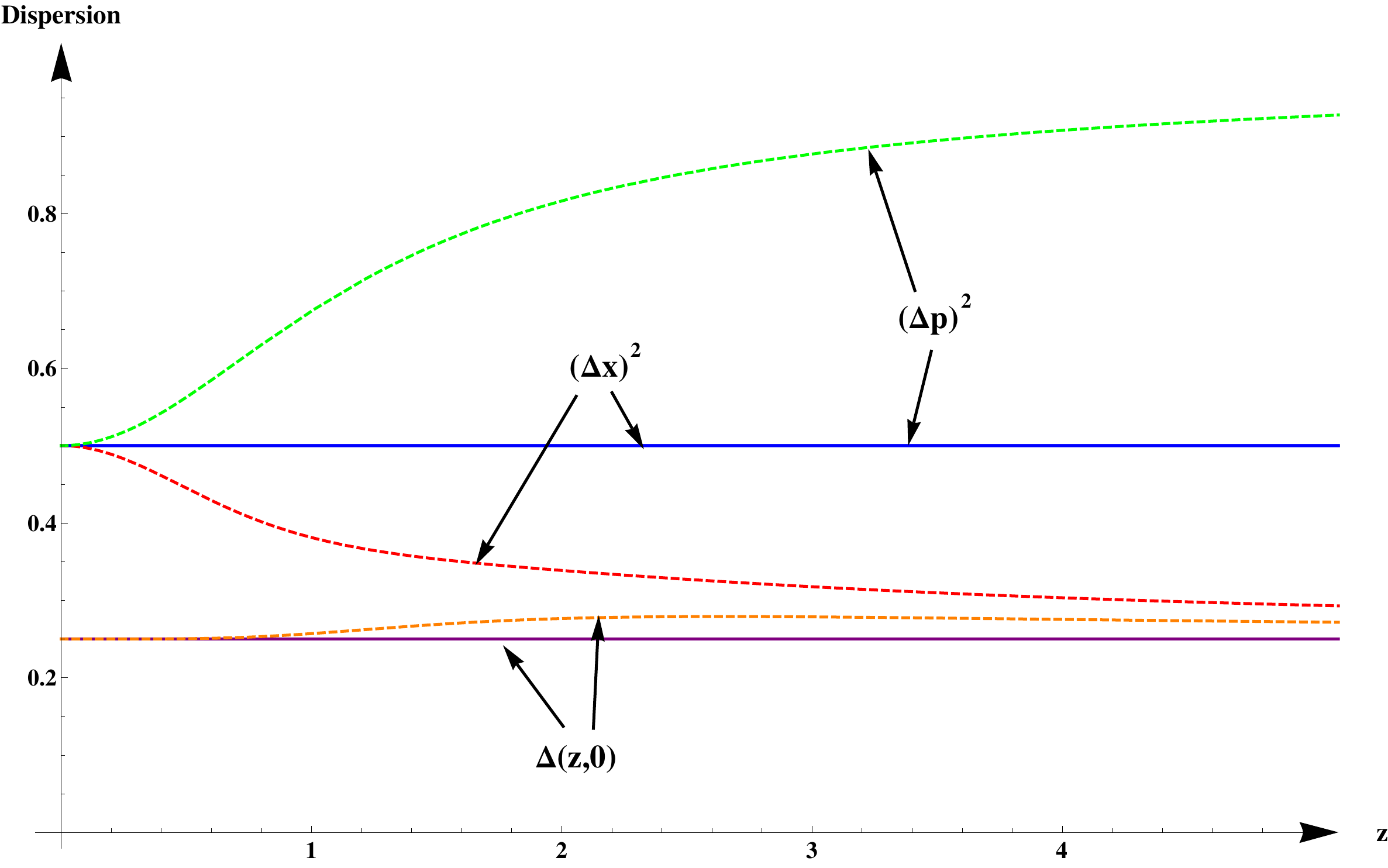}
\caption{ {\small Dispersion in position $(\Delta x)^2$,  momentum $(\Delta p)^2$  and product of dispersions $\Delta (z,0)$ for the usual (plain line) and quadratic (dashed line) CS ($\gamma=0$) of the harmonic oscillator.}}
\label{DispersionOscillateur}
\end{figure}

In our calculations, we choose units to be such that $\hbar=2m=1$ and $\omega=2$. This particular case is interesting since,  for the usual CS, we have $(\Delta x)^2=(\Delta p)^2=1/2$. The mean values and dispersion of $x$ and $p$ are calculated in the two types of CS ($\gamma=0$) and we have the following observations. As we could have expected, the CS constructed with $f(n)=\sqrt{n}$ are more non-classical than the usual ones. Indeed, we see in Fig \ref{DispersionOscillateur} that the product of the dispersions $\Delta(z,0)=(\Delta x)^2(\Delta p)^2$ at $t=0$, no matter what are the values of the amplitude $z$, is more minimized in the usual CS, reaching as expected, the value $1/4$. Note though that in the quadratic case, it remains close to this value. However, the dispersions in position and momentum show a completely different behaviour. We have, as expected, $(\Delta x)^2=(\Delta p)^2=\frac12$ for the usual CS, but we see that, in the deformed ones,  $(\Delta x)^2$ decreases continuously in $z$ while $(\Delta p)^2$ increases maintaining the product almost the same.

\begin{figure}[h!]
\centering
\includegraphics[trim=0 0 7.5cm 0cm,clip,width=.07\textwidth]{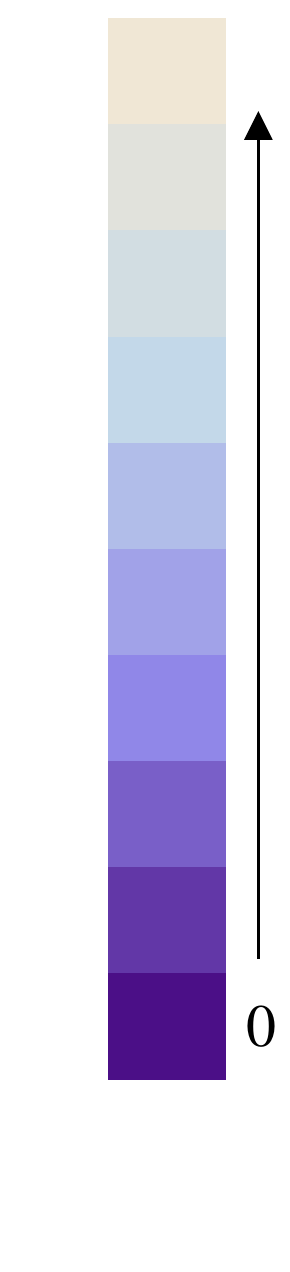} \quad
\includegraphics[width=.35\textwidth]{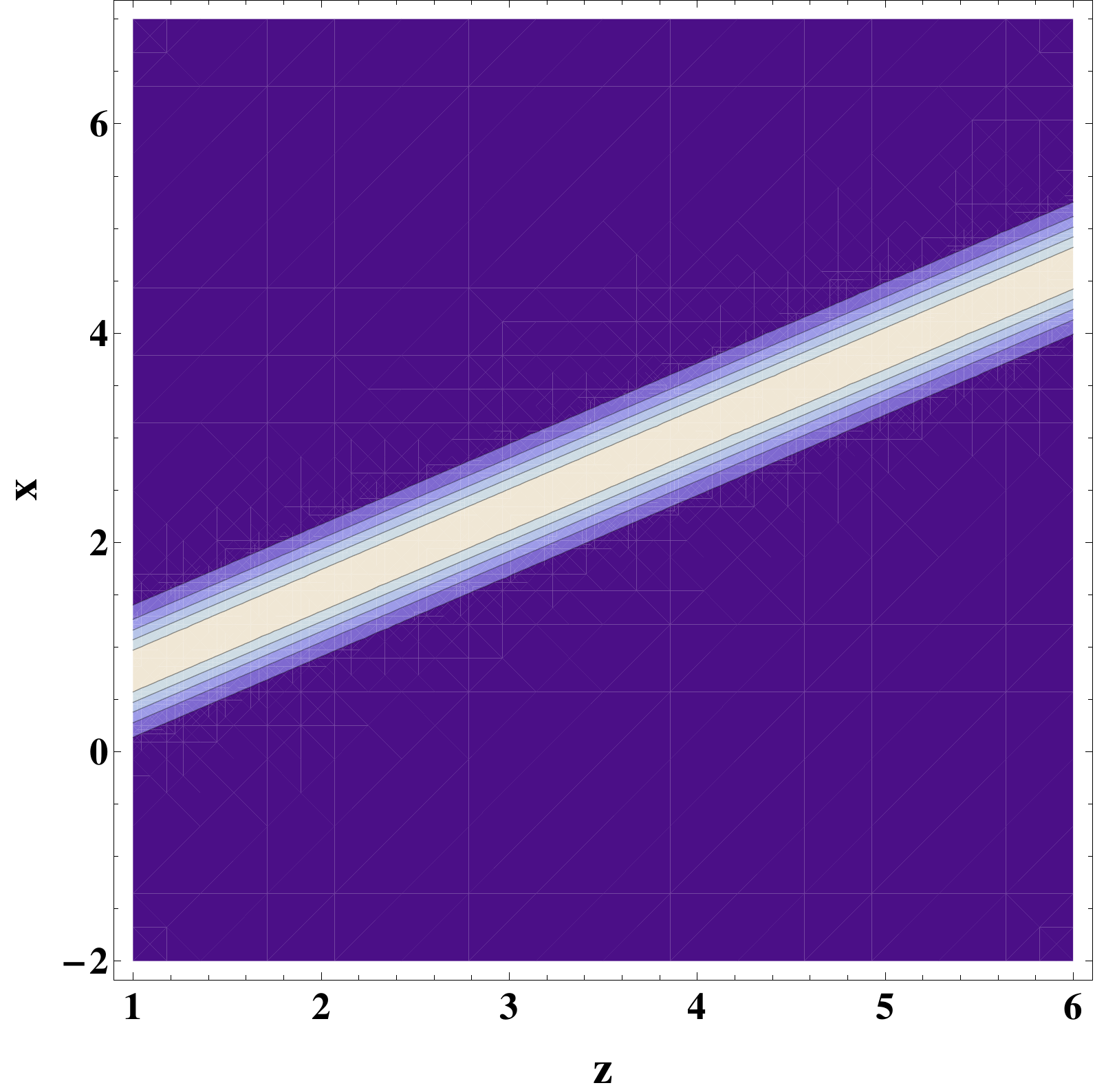} \quad
\includegraphics[width=.35\textwidth]{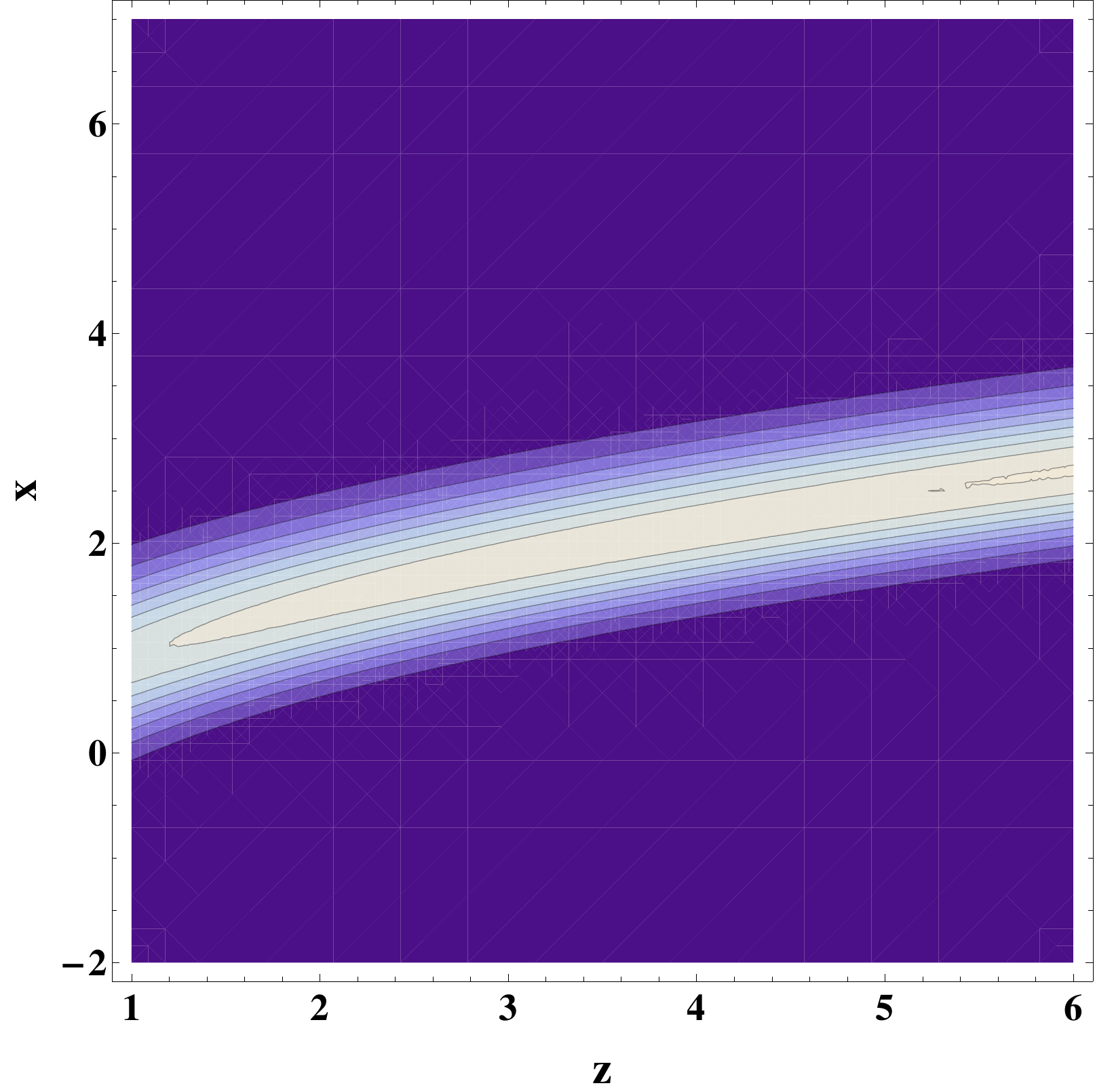} 
\caption{ {\small Comparison of the density probability $|\Psi_{OH}(z,0,x;0)|^2$ for the usual (left) and quadratic (right) CS of the harmonic oscillator.}}
\label{densityOH}
\end{figure}

These results are in agreement with Fig \ref{densityOH} which shows the density probability of both types of CS. For usual CS, we know that changing the value of the amplitude $z$ has the effect of translating the probability density in $x$ without any change of its form and its maximum value. For the quadratic CS, the density probability becomes narrower (and thus higher) when $z$ increases. It is what we expected, since Fig \ref{DispersionOscillateur} shows that the dispersion in $x$ becomes smaller with $z$ increasing. Note that the density probabilities for both types of SCS are similar, but the squeezing is apparent in the quadratic case even when $\gamma=0$.

\begin{figure}[!h]
\centering
\includegraphics[width=.7\textwidth]{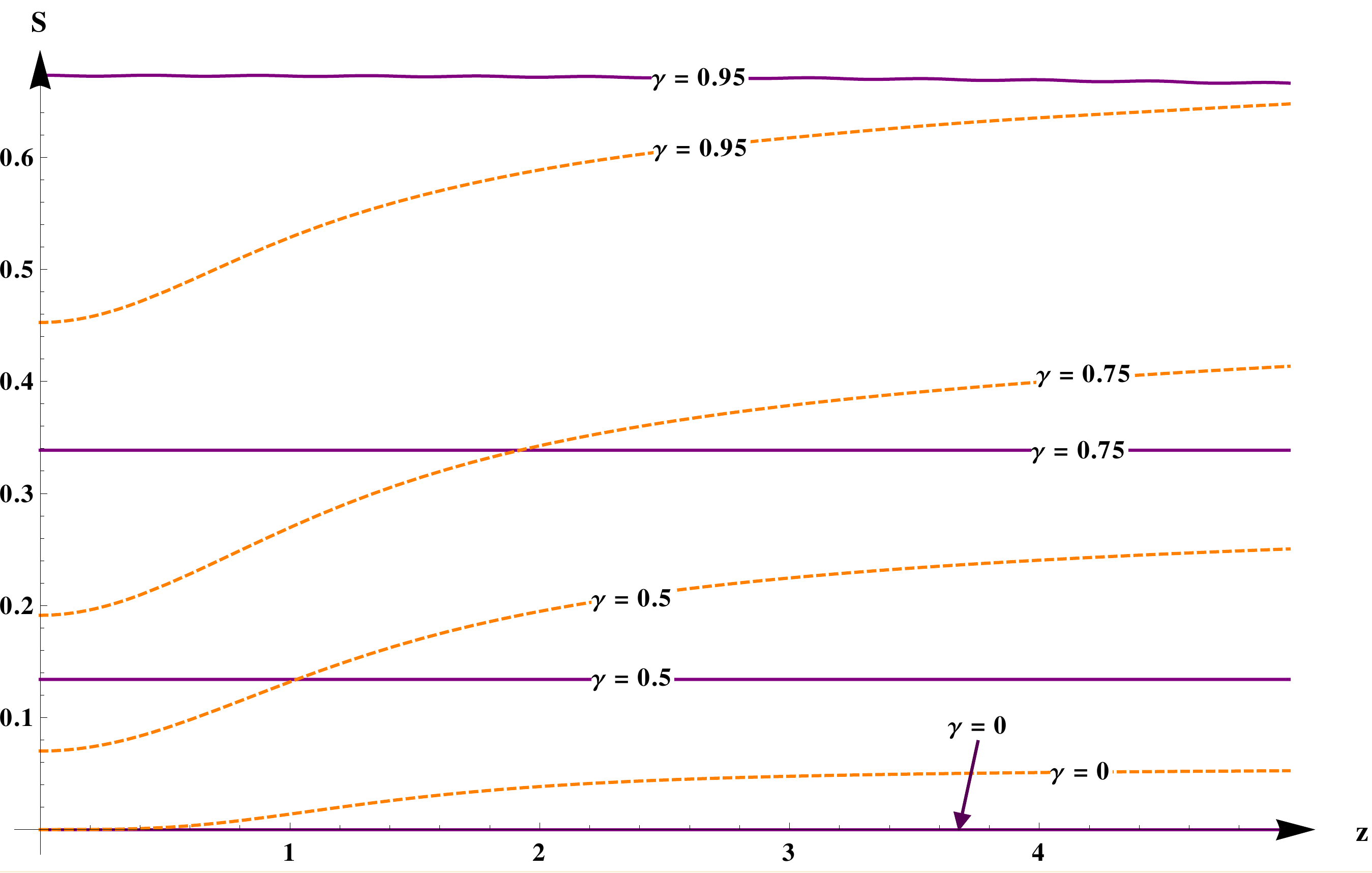}
\caption{ {\small Linear entropy $S$ for the usual (plain line) and quadratic (dashed line) SCS of the harmonic oscillator for different values of the squeezing $\gamma$. }}
\label{ComparaisonEntropyOscillateur}
\end{figure}

The entanglement of the usual SCS  and of some deformed ones has received much attention \cite{Abdel,add,add2}. Let us mention that in these approaches the deformation function $f(n)$ always tends to $1$ at a certain limit. It is not the case for our quadratic CS. Furthermore, the preceding approaches meanly used the von Neumann entropy.  Here we use the linear entropy. As we can see on Fig \ref{ComparaisonEntropyOscillateur},  for small values of $z$ and $\gamma\neq 0$, more entanglement is created with the usual CS. It corresponds to the case when the uncertainty is almost the same for both types of CS. For bigger value of $z$, depending on the value of the squeezing parameter, more entanglement can be created with the quadratic SCS.

In general, if we fix the type of SCS, the maximum entanglement appears, as expected,  when the squeezing parameter $\gamma$ tends to its maximum value close to $1$. We also see, as it is well known, that the usual CS ($\gamma=0$)  have always a null entropy, i.e. no entanglement is created. The quadratic CS, however, create always entanglement.
%


\section{The Morse squeezed coherent states and measure of entanglement }

\subsection{The model}

The one-dimensional Morse potential model is a well known solvable oscillating system composed of two atoms given by the energy eigenvalue equation:
\begin{equation}
{\hat{H}}\ \psi(x) =\left(\frac{{ \hat{p}}^2}{2m_r}+V_M(x)\right)\psi(x)= E\,\psi(x),
 \end {equation}
where the potential is
\begin{equation*}
V_M(x)=V_0(e^{-2\beta x}-2e^{-\beta x}) .
\end{equation*}

\begin{figure}[h!]
\centering
\includegraphics[width=.5\textwidth]{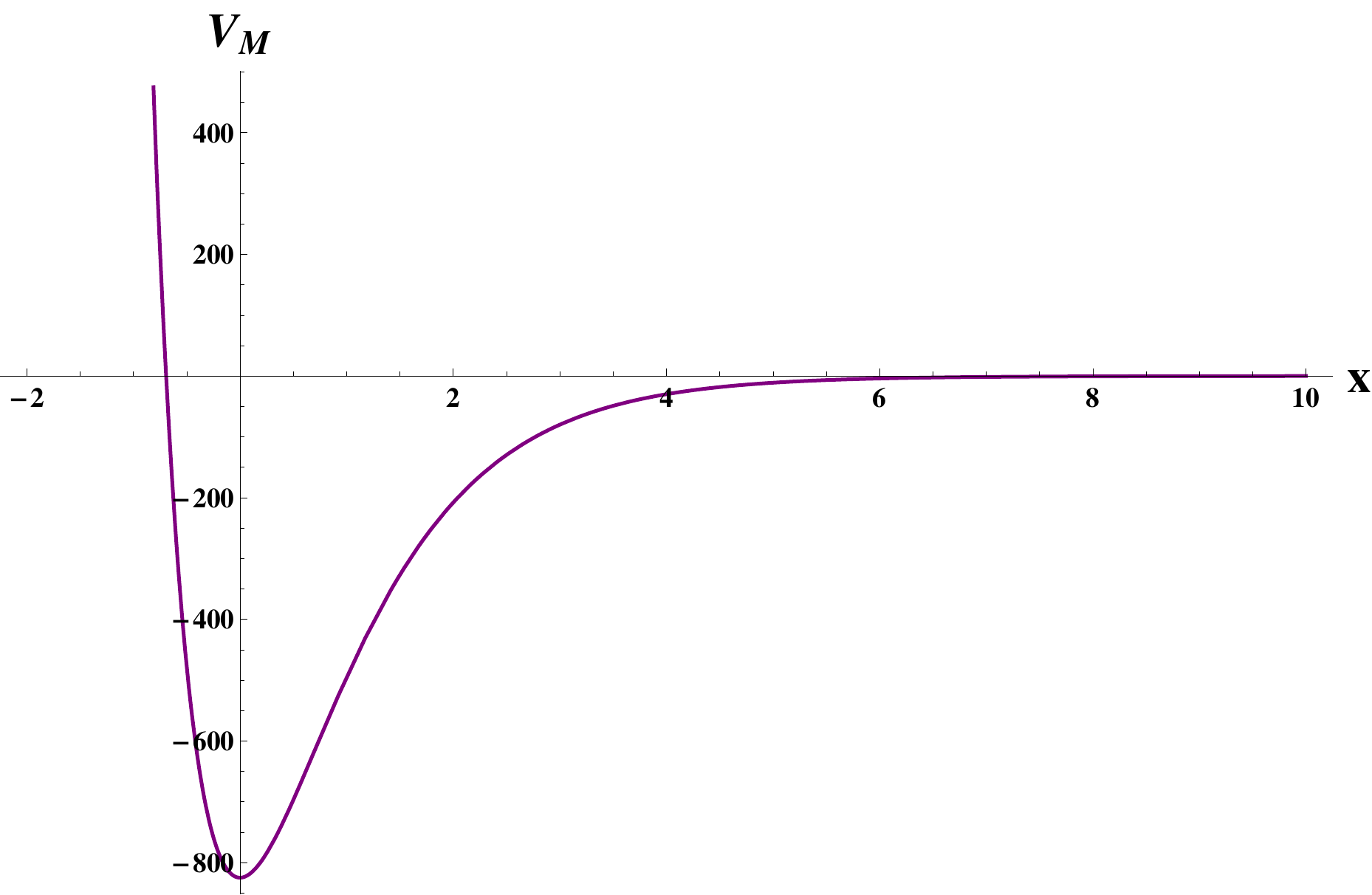}
\caption{ {\small Morse potential for the hydrogen chloride molecule. }}
\label{potentielMorse}
\end{figure}

In this paper, we use the same definitions and notations as in  \cite{angelova12}. The unit convention is $\hbar=2m_r=1$ and $\beta=1$. For all the calculations and graphs, we are considering the example of  the hydrogen chloride molecule for which the number of energy eigenstates is denoted by $[p]$. The parameter $p$ is a physical quantity related to the parameters of the system under consideration and, in our case, $p=28.22$ \cite{AF07}. The finite discrete spectrum can thus be written as
\begin{equation}
E_n=-(p-n)^2
\label{energies}
\end{equation}
and the corresponding energy eigenfunctions are
\begin{equation}
\psi_{n, \rm{Morse}}(x)= {\cal N}_n \ e^{-\frac{y}{2}} y^{\epsilon_n} L_{n}^{2\epsilon_n}(y),
\label{psinMorse}
\end{equation}
where we have used the change of variable $y=(2p+1) e^{-x}$. The functions $L_{n}^{2\epsilon_n}(y)$ are the associated Laguerre polynomials and 
${\cal N}_n=\sqrt{ \frac{2(p-n)\Gamma(n+1)}{\Gamma(2p-n+1)}}$ is the normalization factor.

Two different types of SCS have been considered in \cite{angelova12}. They are given by (\ref{scs}) where $M=[ p ]-1$ and the $|n\rangle$  are energy eigenstates (\ref{psinMorse}). Let us explicitly write those states, evolving in time, as:
\begin{equation}
\Psi_{\rm{Morse}}(z, \gamma, x;t)=  \frac{1}{\sqrt{{\cal{N}} (z,\gamma)}}
\sum_{n=0}^{M} \frac{Z(z,\gamma,n)}{\sqrt{n!}f(n)!} e^{-i\frac{E_n}{\hbar}t} \psi_{n, \rm{Morse}}(x).
\label{scsMorsetime}
\end{equation}
The first type is given by the oscillator-like SCS  since $f(n)=1$. They are similar to the ones for the harmonic oscillator (undeformed) except that the superposition of eigenstates  is now finite. The second type is called energy-like SCS since the product of the ladder operators factorizes the Hamiltonian. They are such that $f(n)=\sqrt{2p-n}$ and we have $Z(z,\gamma,n)=Z_e^{2p} (z,\gamma,n)$  for $n=1,2,...,[p]-1 $ given by 
\begin{equation}
Z_{e}^{2p}(z,\gamma,n)=(-1)^n \gamma^{\frac{n}{2}} \frac{\Gamma(2p)}{\Gamma(2p-n)}\ {}_2 F_1\left(\begin{matrix}-n, - {\frac{z}{2 \sqrt\gamma}}+{\frac{1-2p}{2}}  \\ 1 - 2p\end{matrix};2\right).
\label{Zenergy}
\end{equation} 

Let us summarize the results and observations given in \cite{angelova12} about the behaviour of those states. The phase space trajectories for both types of SCS show always a squeezing effect even if $\gamma=0$, i.e., they are elliptic (with large eccentricity). The  oscillator-like SCS are less stable as time evolves than the energy-like SCS for the same value of $z$ as we see from the phase space trajectories (see in  \cite{angelova12}). When $z<20$ in our units, the energy-like SCS are mostly minimum uncertainty states and are well localized in position. For the  oscillator-like SCS, we observe similar results but the values of $z$ are smaller ($z<3$).

Our main observation was thus that  energy-like SCS  are less non classical than the oscillator-like one and our concern now is to confirm this fact by computing the level of entanglement of our SCS states using the linear entropy.  

\begin{figure}[h]
\centering
\includegraphics[width=.7\textwidth]{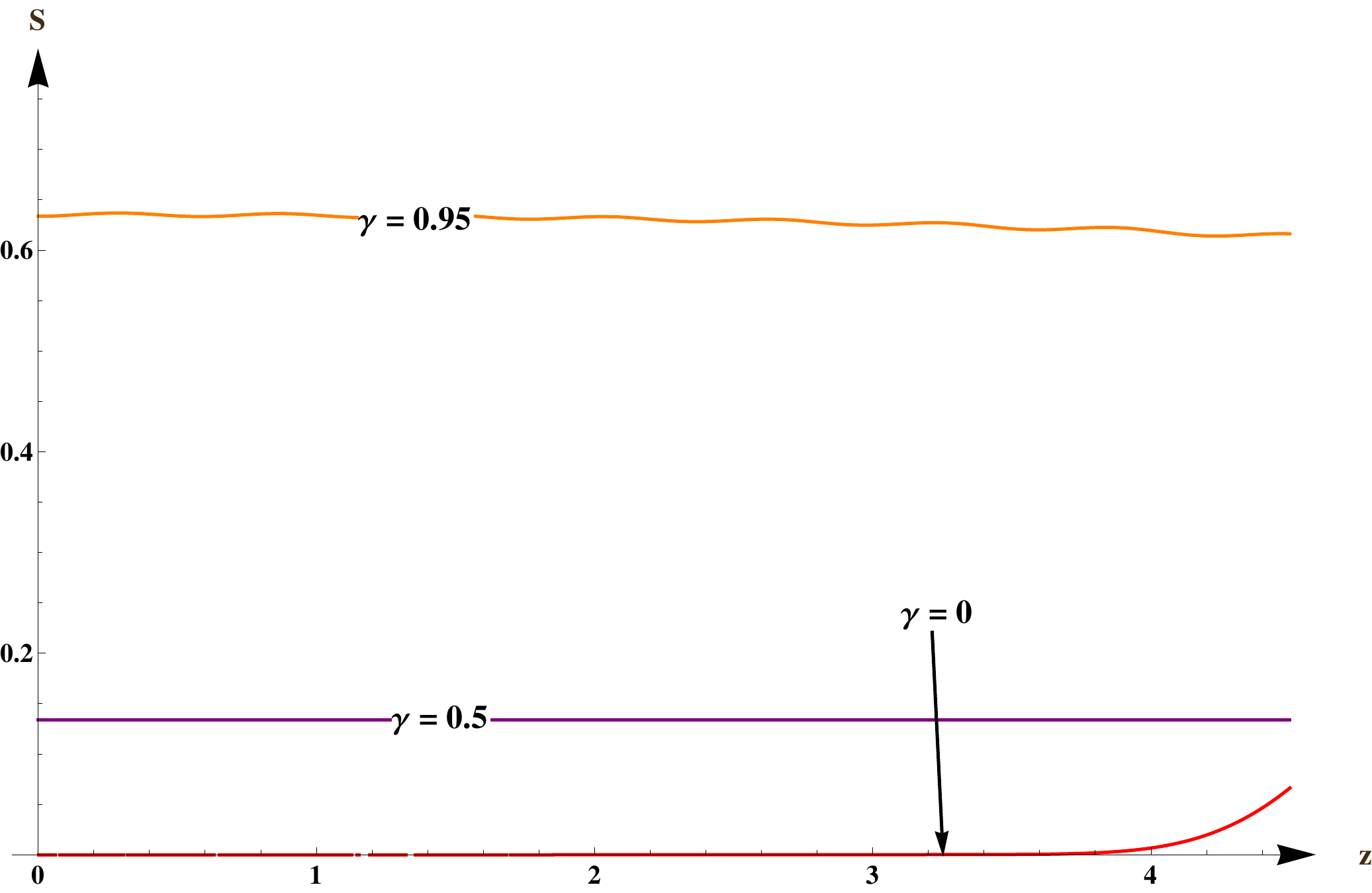}
\caption{ {\small Linear entropy for the Morse potential  oscillator-like  SCS as a function of $z$ for different values of the squeezing $\gamma$.}}
\label{CSCMorseOscillator}
\end{figure}

For the oscillator-like SCS, we see that the linear entropy is essentially the same as for the usual harmonic oscillator since the equation (\ref{entropy}) does not depend on the eigenstates of the system. The only difference is the number of states in the sums. 
Fig \ref{CSCMorseOscillator} gives the linear entropy for those states as a function of $z$ for different values of $\gamma$ where we have chosen $z$ and $\gamma$ real with the same range of values as in \cite{angelova12}. Again, we see that adding squeezing in our states has as an effect to create more entanglement. Let us insist on the fact that entanglement is not very sensitive to the value of $z$ (at least when $z$ is small enough).

\begin{figure}[h!]
\centering
\includegraphics[width=.7\textwidth]{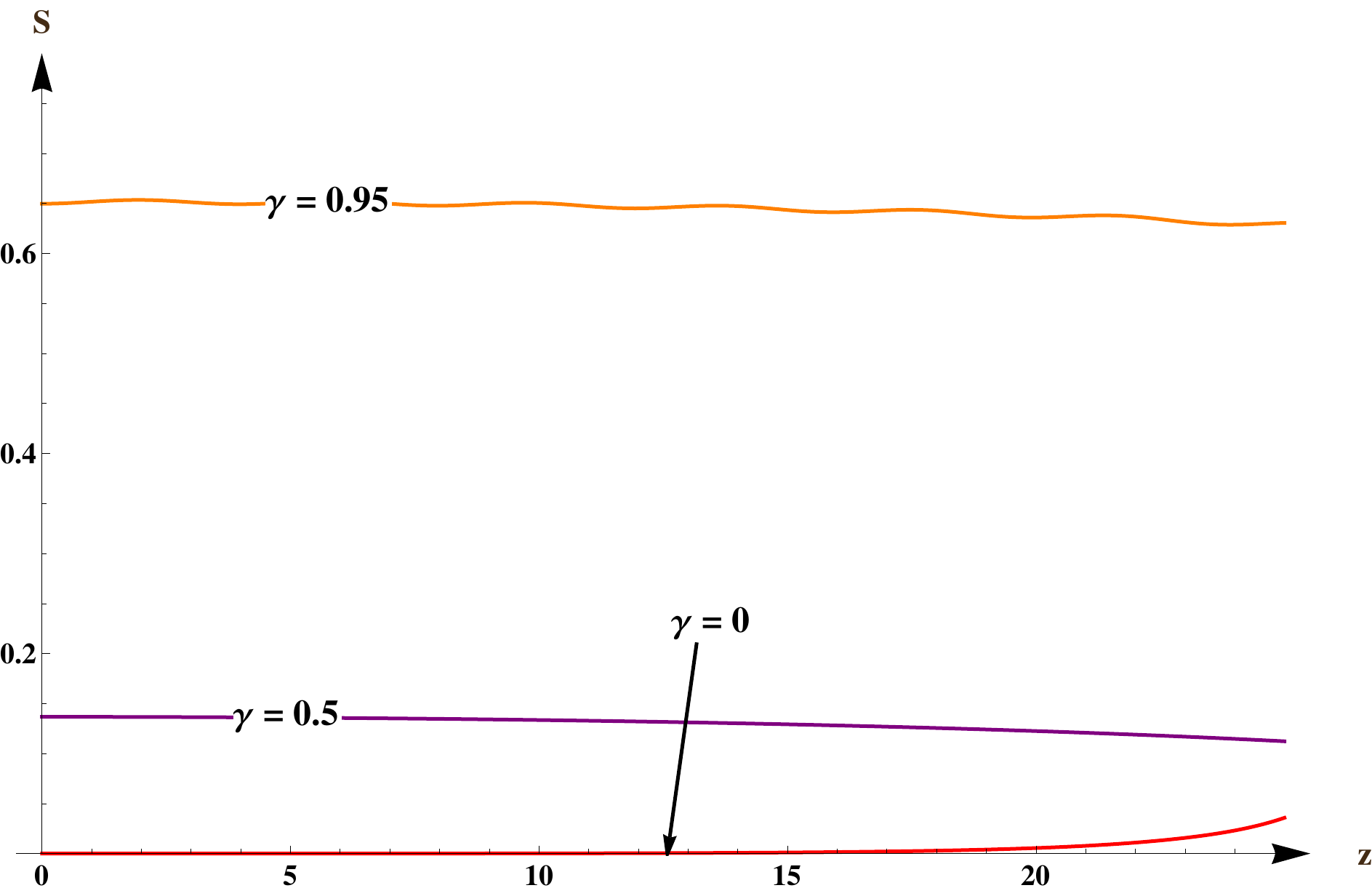}
\caption{ {\small Linear entropy for the Morse potential  energy-like  SCS as a function of $z$ for different values of the squeezing $\gamma$.}}
\label{CSCMorseEnergy}
\end{figure}

Fig \ref{CSCMorseEnergy} shows a similar behaviour of the linear entropy for the energy-like SCS but now we have taken a bigger range of values of $z$ as in \cite{angelova12}.

\begin{figure}[h!]
\centering
\includegraphics[width=.35\textwidth]{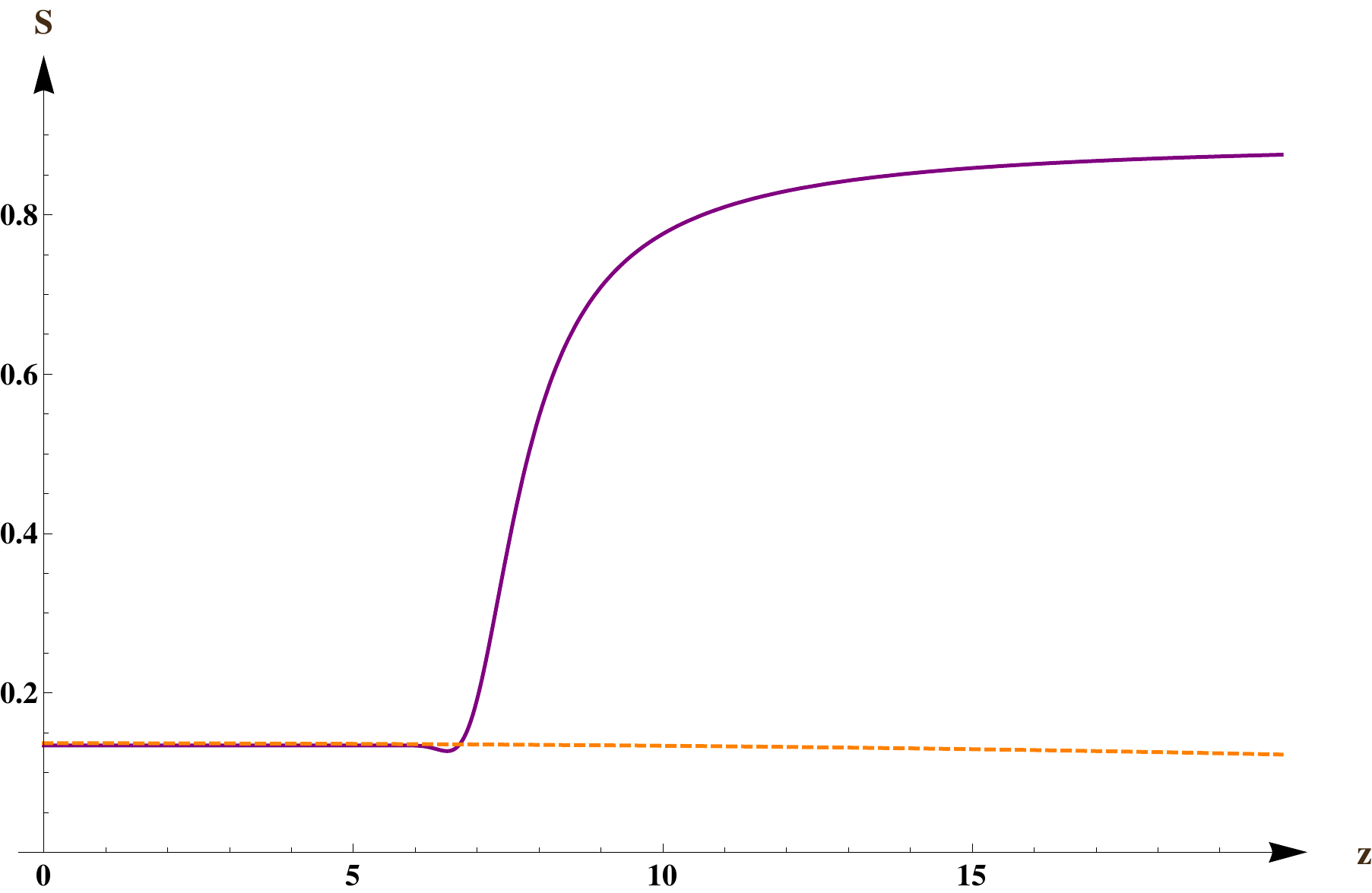} \qquad \qquad
\includegraphics[width=.4\textwidth]{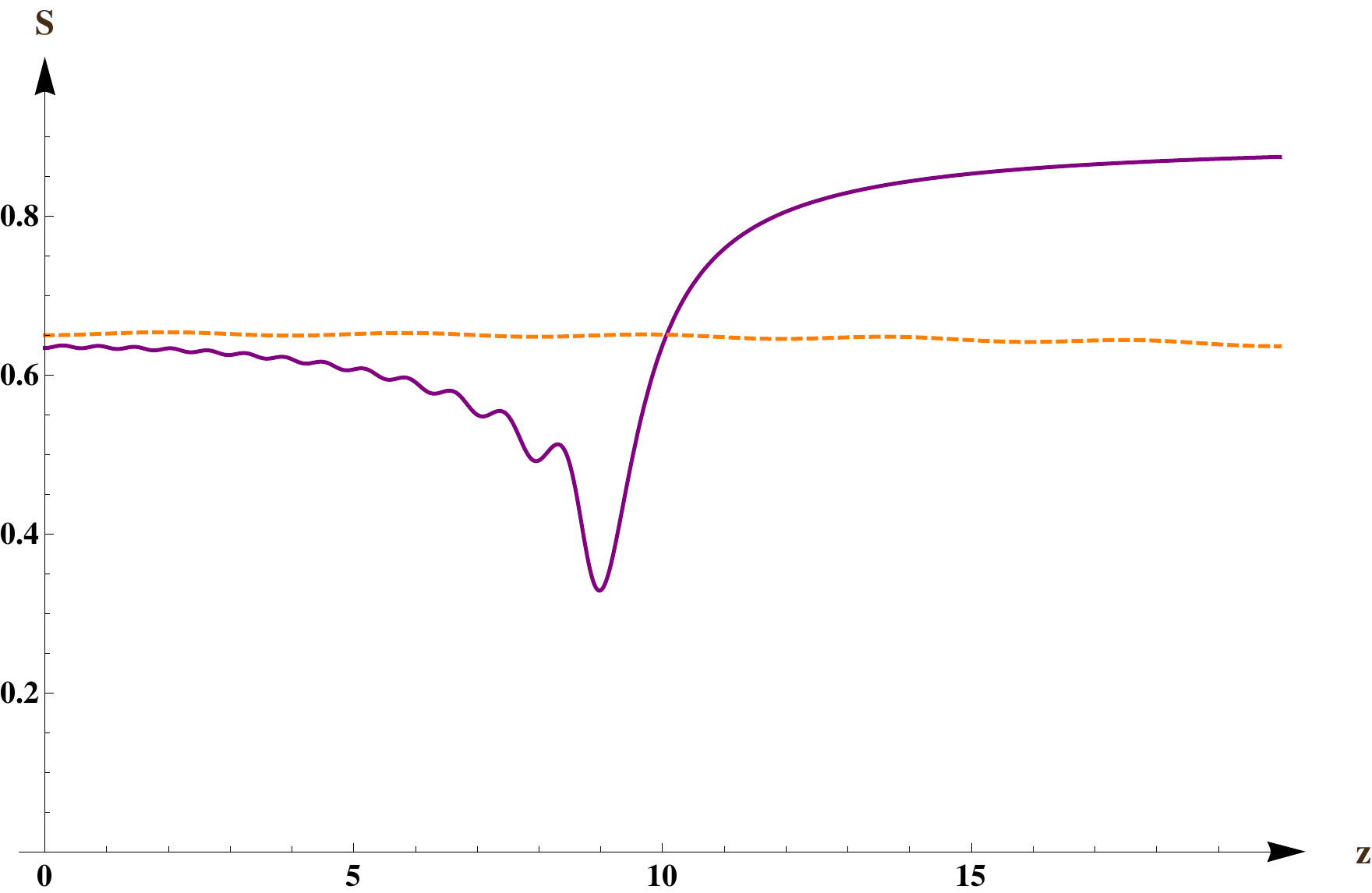}
\caption{ {\small Comparison of the linear entropy for  oscillator-like (plain line) and energy-like (dashed line) SCS as a function of $z$ with moderated squeezing $\gamma=0.5$ (left) and big squeezing $\gamma=0.9$ (right).}}
\label{ComparaisonEnergyOscillator}
\end{figure}

\begin{figure}[h!]
\centering
\includegraphics[trim=0 0 7.5cm 0cm,clip,width=.07\textwidth]{Figure2a} \ 
\includegraphics[width=.35\textwidth]{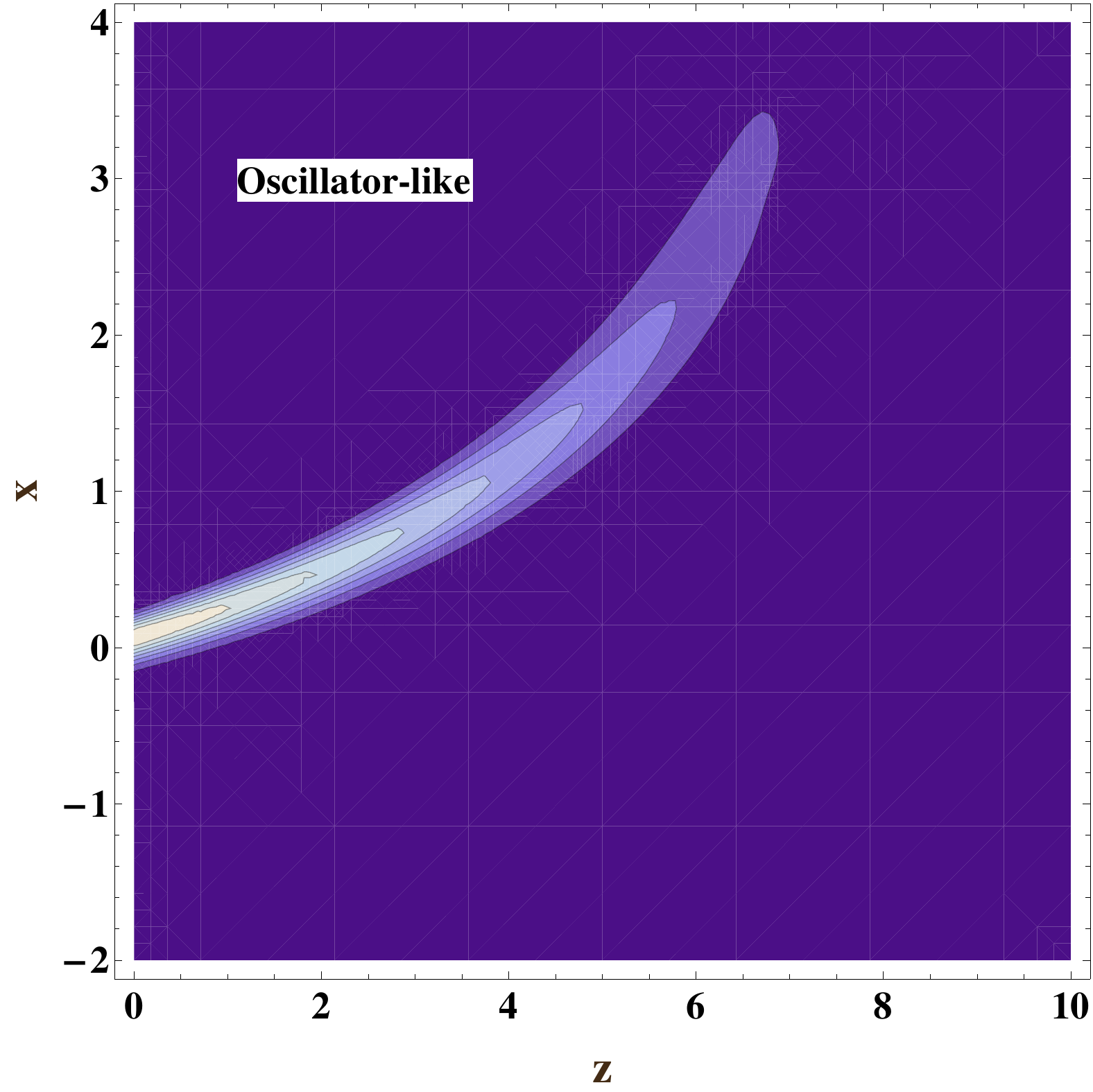} \ 
\includegraphics[width=.35\textwidth]{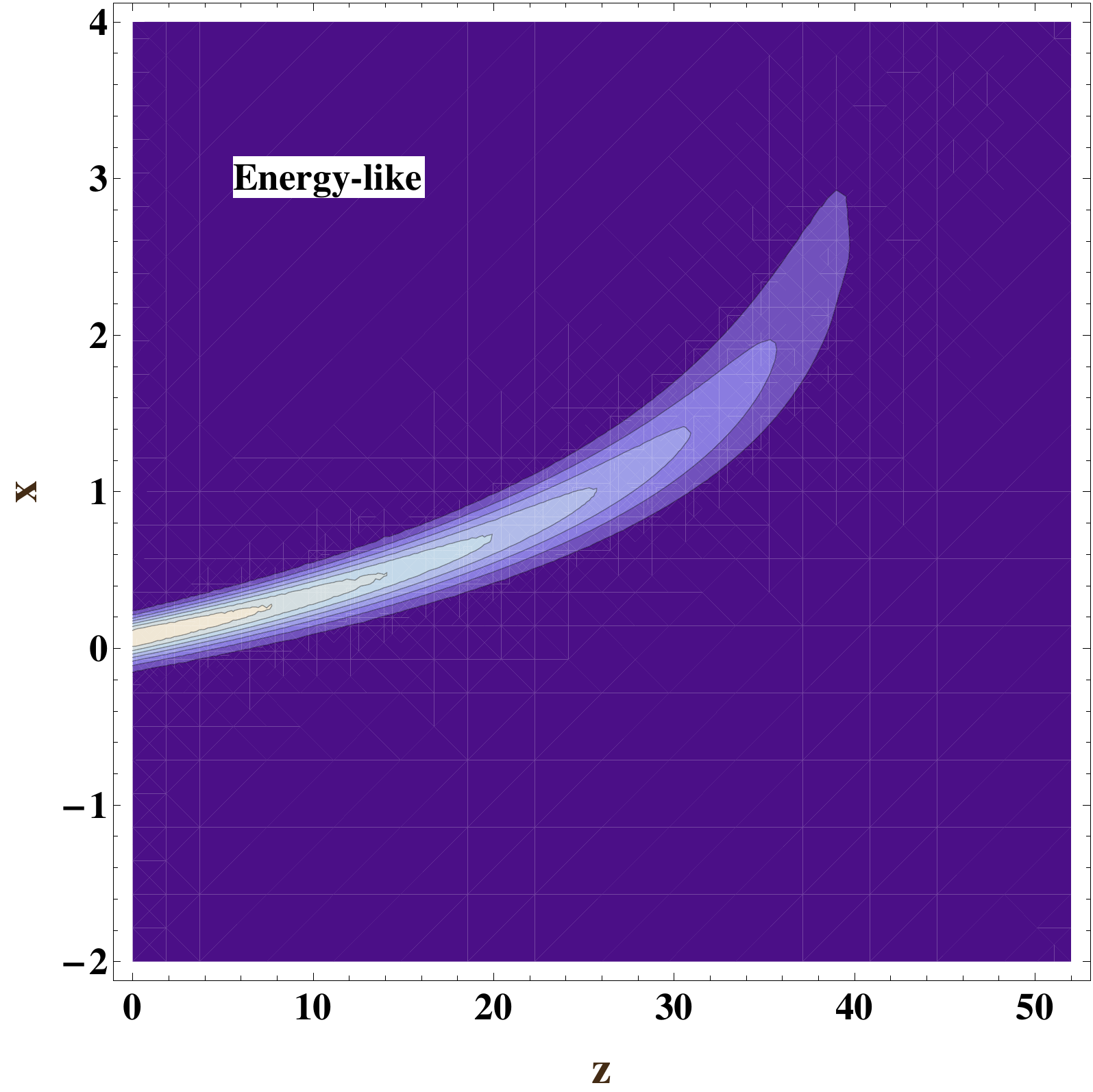}
\caption{ {\small Comparison of the density probability $|\Psi_{Morse}(z,0.5,x;0)|^2$ for the oscillator-like and energy-like SCS of the Morse potential.}}
\label{density}
\end{figure}

In Fig \ref{ComparaisonEnergyOscillator} , we compare the linear entropy created by both types of SCS.  If we look at the graphs for amplitude $z$ smaller than $5$, we see that both types create almost the same quantity of entanglement for a fixed squeezing $\gamma$, but the oscillator-like SCS create a little less entanglement than the energy-like SCS. However, for the oscillator-like SCS, the behaviour completely changes for $z>5$ creating big entanglement. It means that the energy-like SCS are more classical for a bigger range of values of the amplitude $z$. This result is clearly related to the behaviour of our states with respect to the density probability. First let us mention that it is really different than in the harmonic case due to the form of the potential where, as it can be seen in Fig \ref{potentielMorse}, if $x$ is bigger than approximately 4, the Morse potential tends to 0 and the probability of finding a bound state is almost null.  Second, for both types of SCS, we observe that changing the value of $z$ does not only translate the density probability, but also flattens it. The limit value of $z$ (different for the two types of SCS) observed on the previous figures corresponds to the place where the density probability tends to 0 (see Fig \ref{density}). 


\subsection{Dynamical behaviour of the linear entropy}

Using the time dependent SCS given in (\ref{scstime}), we get the time evolution of the linear entropy as:
\begin{eqnarray}
S(t) = 1 &- &\frac{1}{\mathcal{N}^2}\sum_{q=0}^{M}\sum_{j=0}^{M}\sum_{m=0}^{M-max(q,j)}\sum_{n=0}^{M-max(q,j)} e^{-i(E_{m+q}-E_{m+j}+E_{n+j}-E_{n+q})t}\nonumber\\
&\times\,& \frac{Z(z,\gamma,m+q)\overline{Z(z,\gamma,m+j)}Z(z,\gamma,n+j)\overline{Z(z,\gamma,n+q)}}{q!j!m!n!f(m+q)!f(m+j)!f(n+j)!f(n+q)!} \nonumber\\
 &\times\,&  |t|^{2(q+j)} |r|^{2(m+n)}.
\label{entropyTime}
\end{eqnarray}
We know that, for the harmonic oscillator, the linear entropy is time independent. It is not the case here. Indeed, using the expression (\ref{energies}) of the Morse potential energy $E_n$,  we get
 \begin{equation}
E_{m+q}-E_{m+j}+E_{n+j}-E_{n+q}=-2(m-n)(q-j).
\end{equation}

\begin{figure}[h!]
\centering
\includegraphics[width=.4\textwidth]{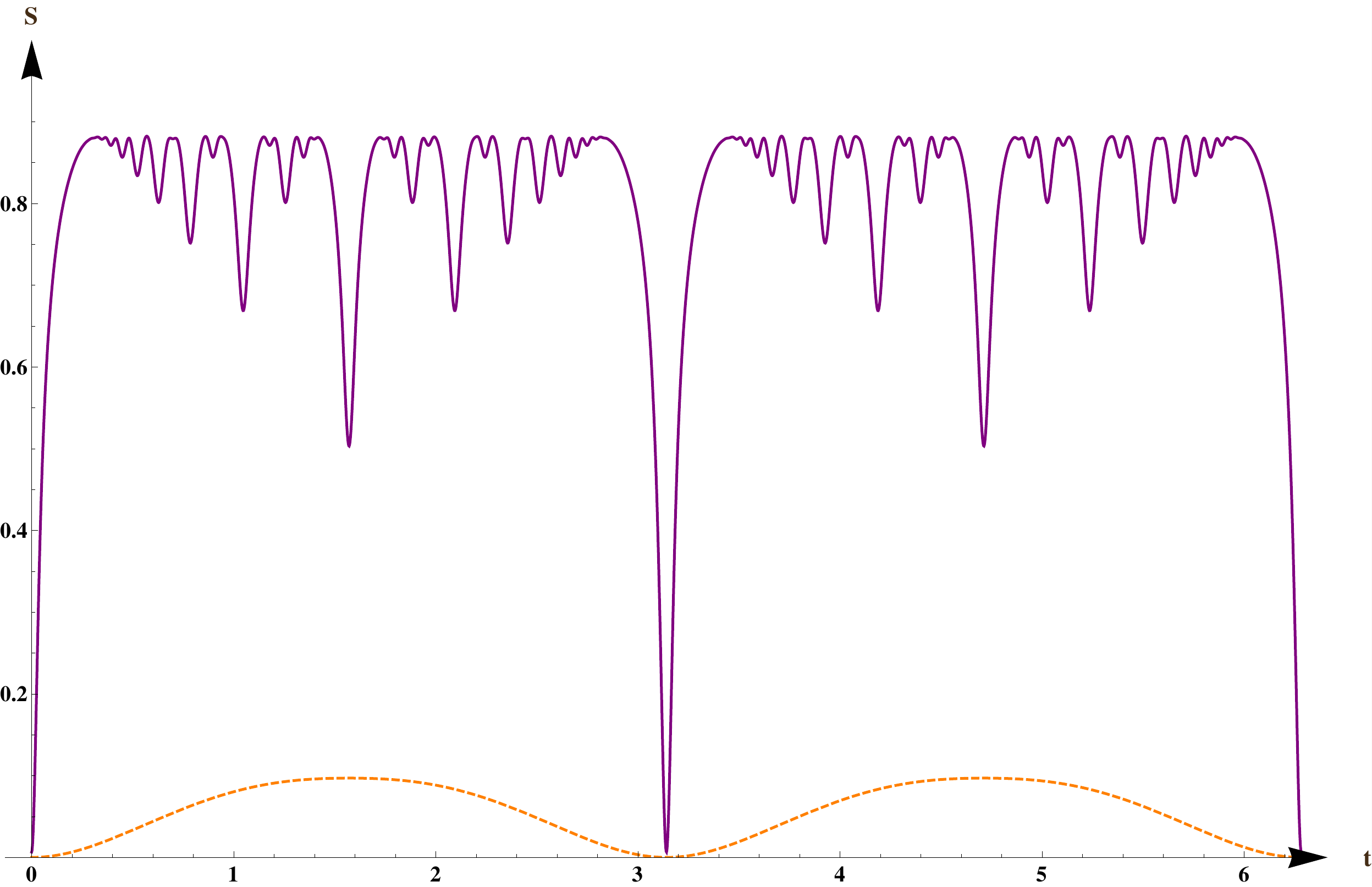} \qquad\qquad
\includegraphics[width=.4\textwidth]{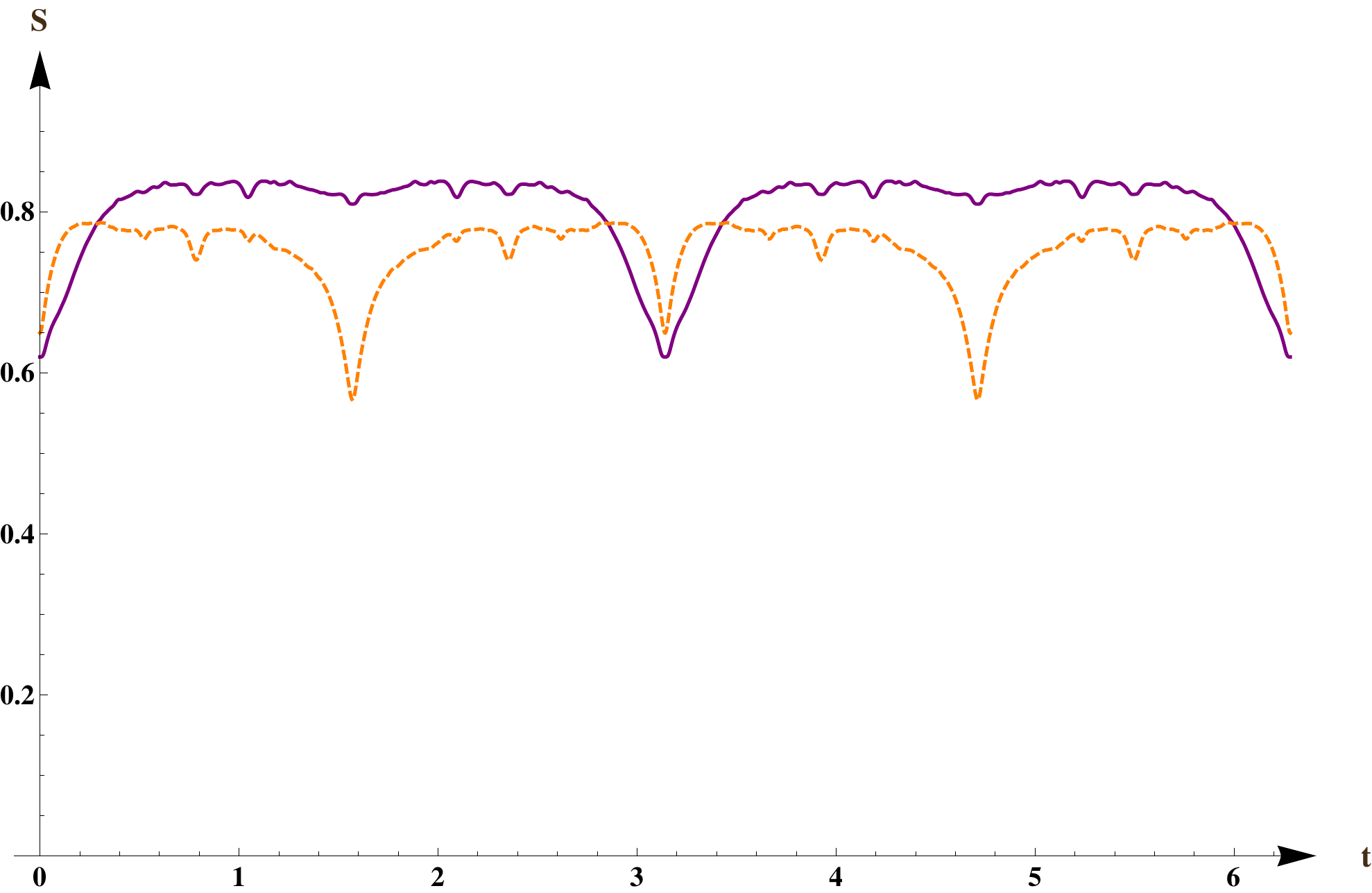}
\caption{ {\small Comparison of $S(t)$ for  oscillator-like  (plane line) and energy-like (dashed line) SCS with an amplitude $z=4$ and without squeezing $\gamma=0$ (left) or with a big squeezing $\gamma=0.95$ (right).}}
\label{TimeEntropySCS}
\end{figure}

\begin{figure}[h!]
\centering
\begin{tabular}{cc}
\includegraphics[width=.4\textwidth]{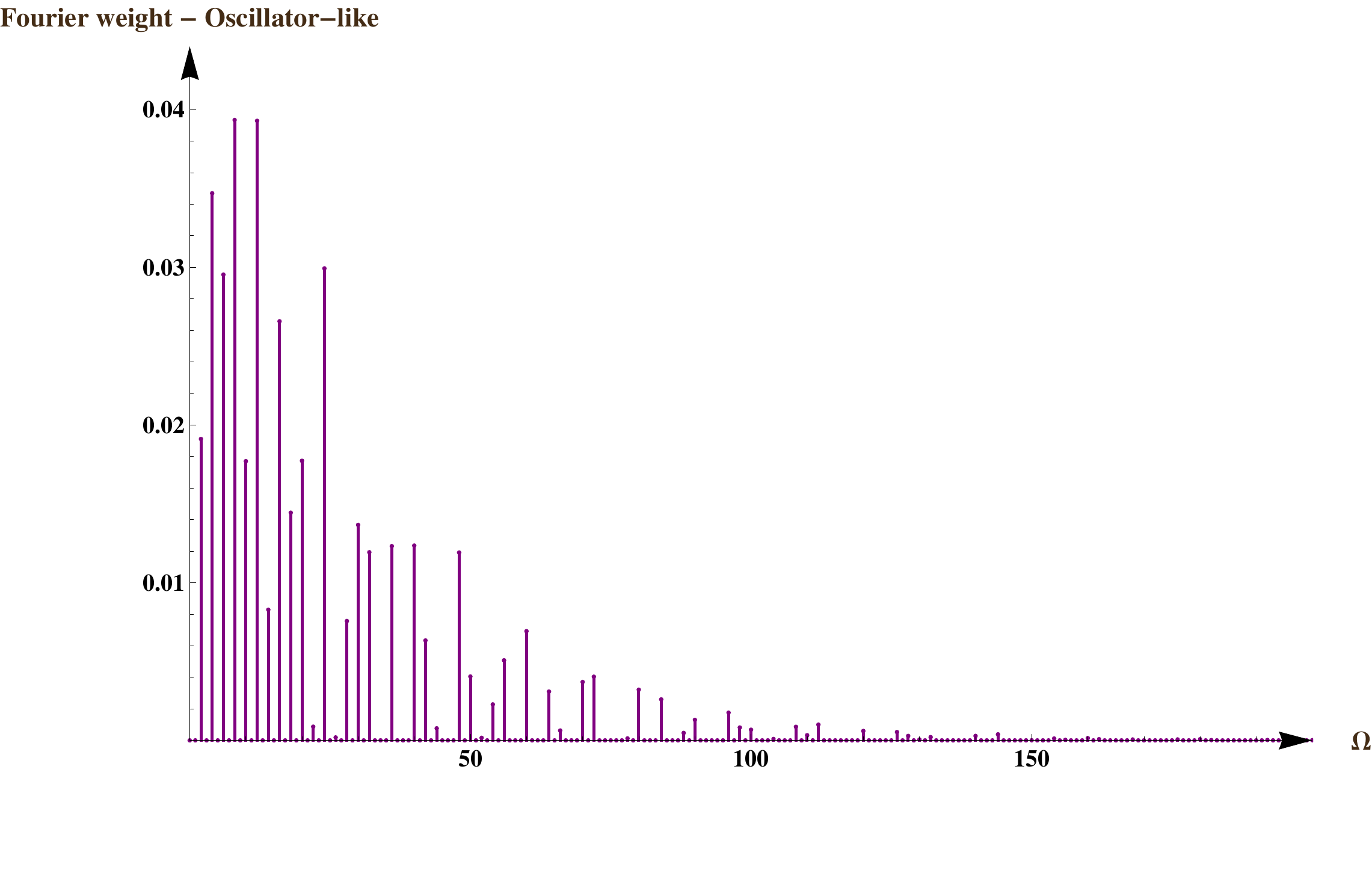} & \includegraphics[width=.4\textwidth]{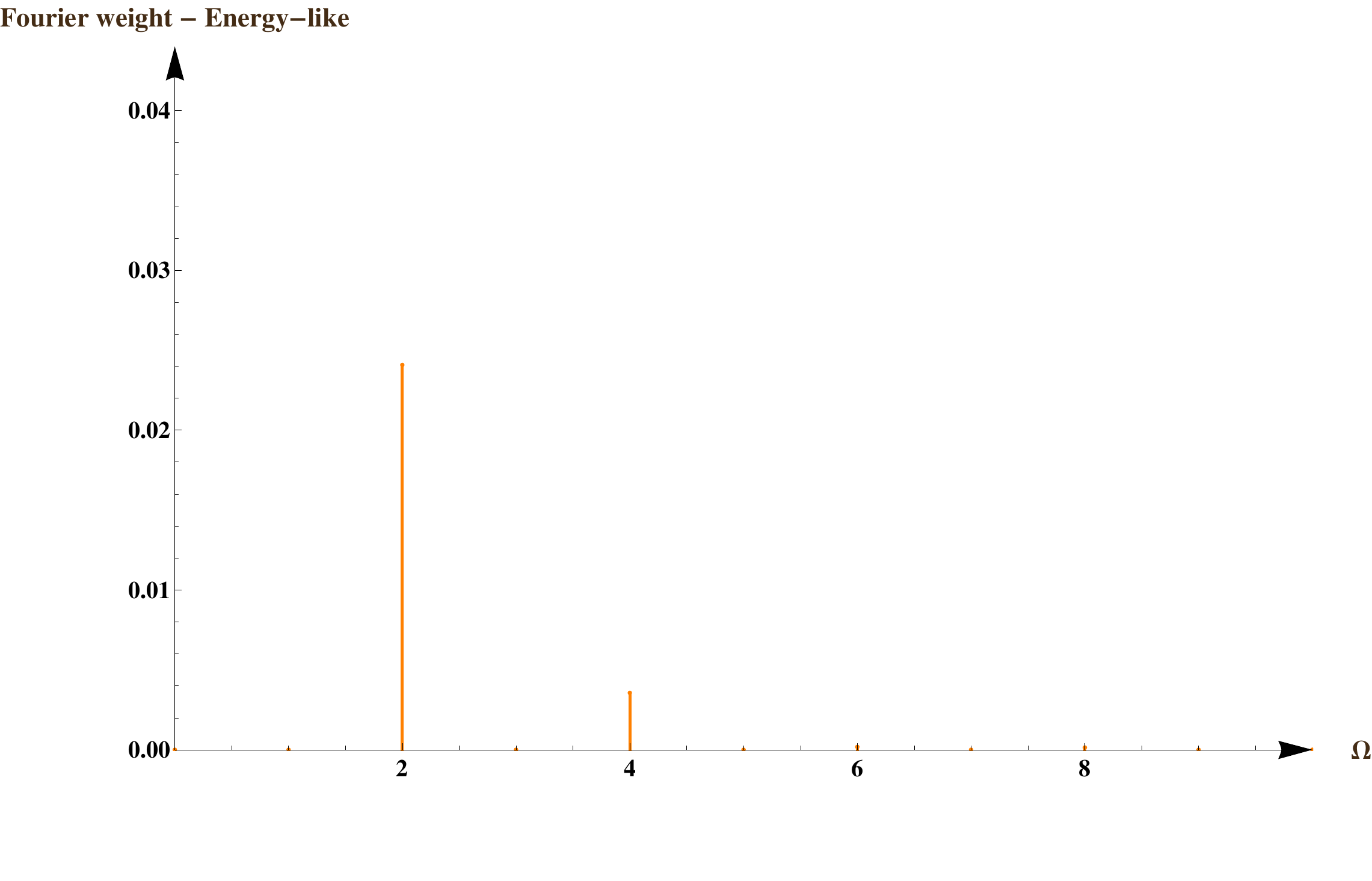}  \\
\includegraphics[width=.4\textwidth]{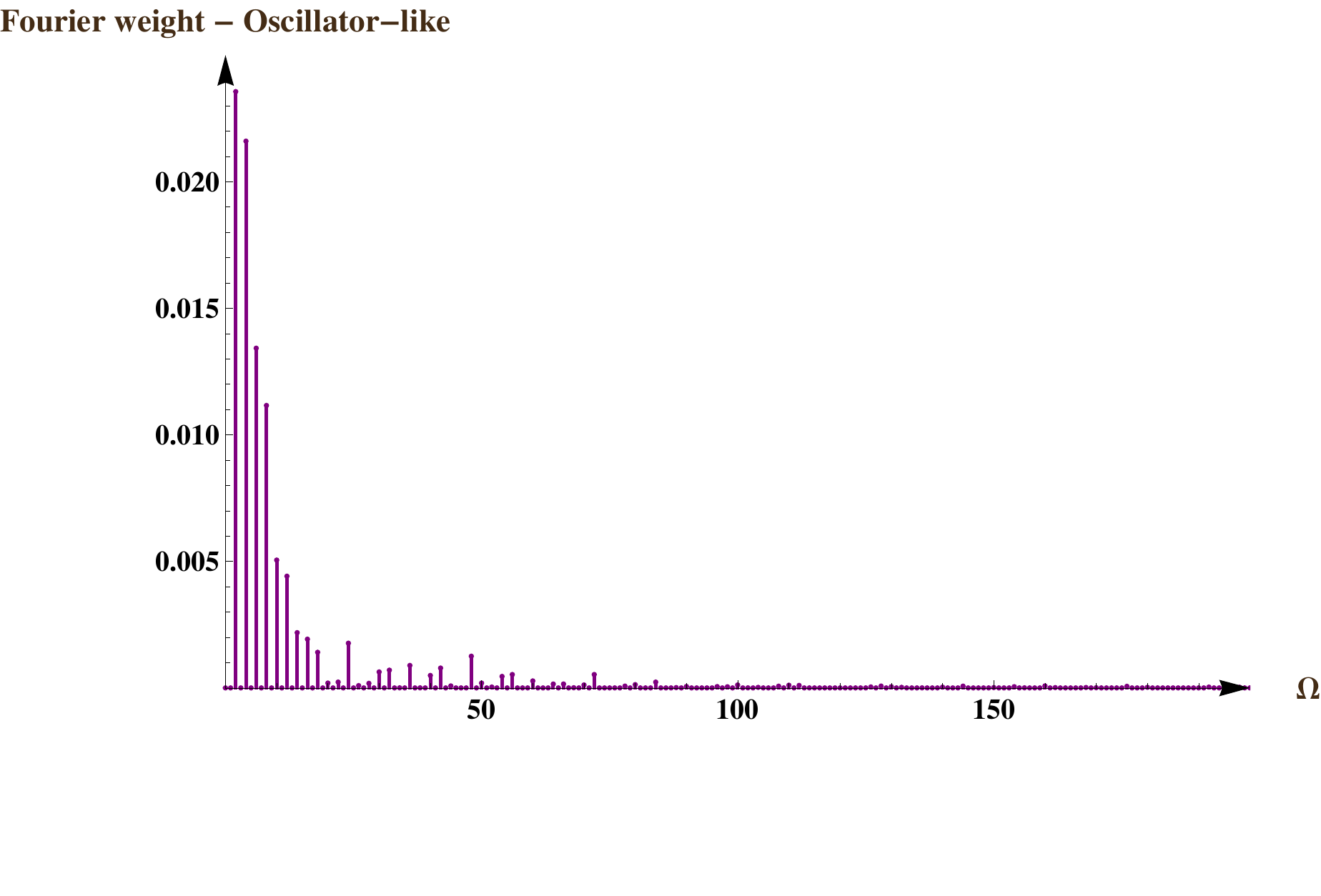} & \includegraphics[width=.4\textwidth]{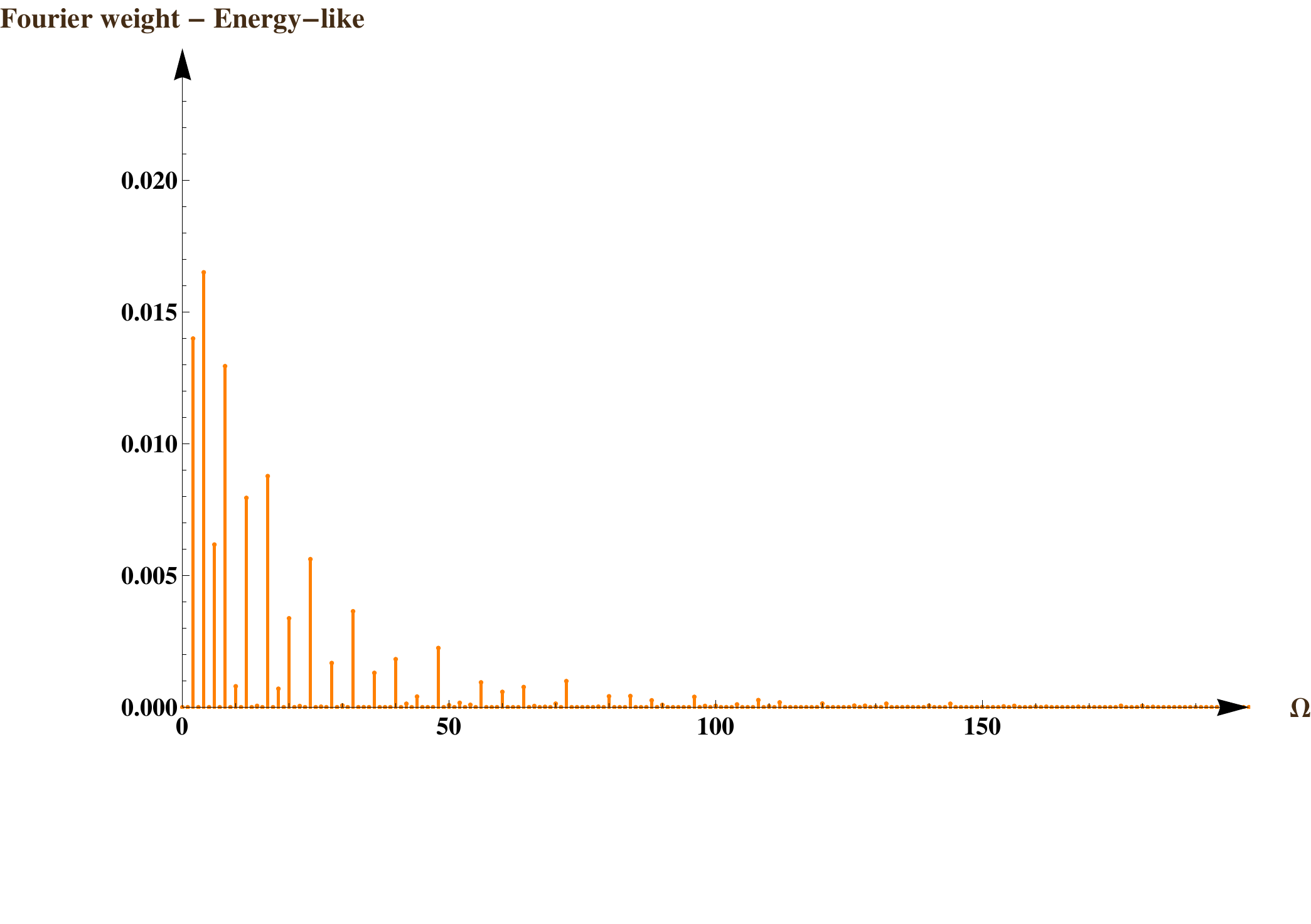}\\
\end{tabular}
\caption{ {\small Fourier transform (with the mean value removed) for  oscillator-like  and energy-like SCS with amplitude $z=4$ and without squeezing, $\gamma=0$ (up), or with a big squeezing, $\gamma=0.95$ (down). We show only positive values of the frequencies since the graph is symmetric.}}
\label{Fourier}
\end{figure}

Thus $S(t)$ is periodic with period $\pi$. On Fig \ref{TimeEntropySCS}, we see that the minimum value of $S$ is reached at $t=0$. Thus, if the aim is to create more entanglement, it is better to let the states evolve in time and select the time for which $S(t)$ is maximum. However, if we are looking for quasi-classisity, i.e., less entanglement, then it is enough to work at $t=0$. We also observe that the two types of SCS do not have the same oscillations. To further analyze these oscillations, we compute in  Fig \ref{Fourier} the Fourier transform of the linear entropy, i.e., we compute the coefficients $C_{\Omega}$ defined as 
\begin{equation}
C_{\Omega}=\frac{1}{\pi}\int^{\pi/2}_{-\pi/2}S(t)\,\,e^{-2\Omega it}dt \,\,\,\,\,\,\,\,\,\,\,\,\,\,\, S(t)=\sum^{\infty}_{\Omega=-\infty}C_{\Omega}e^{2\Omega it}.
\end{equation}

Note that in the plotting, we have removed the mean value of the oscillations (the coefficient $C_{0}$) in order to see better the frequencies amplitudes. Then, when looking at Fig \ref{Fourier}, we see that the first harmonics are more important than the others (here, the fundamental harmonic corresponds to $\Omega=2$). Also, the coefficients linked to different frequencies are not the same, depending on the type of SCS. This is what explains that the two plots have not the same amplitude of oscillations.

\subsection{Entanglement and non-symmetric beam splitter}

\begin{figure}[h]
\centering
\includegraphics[width=.6\textwidth]{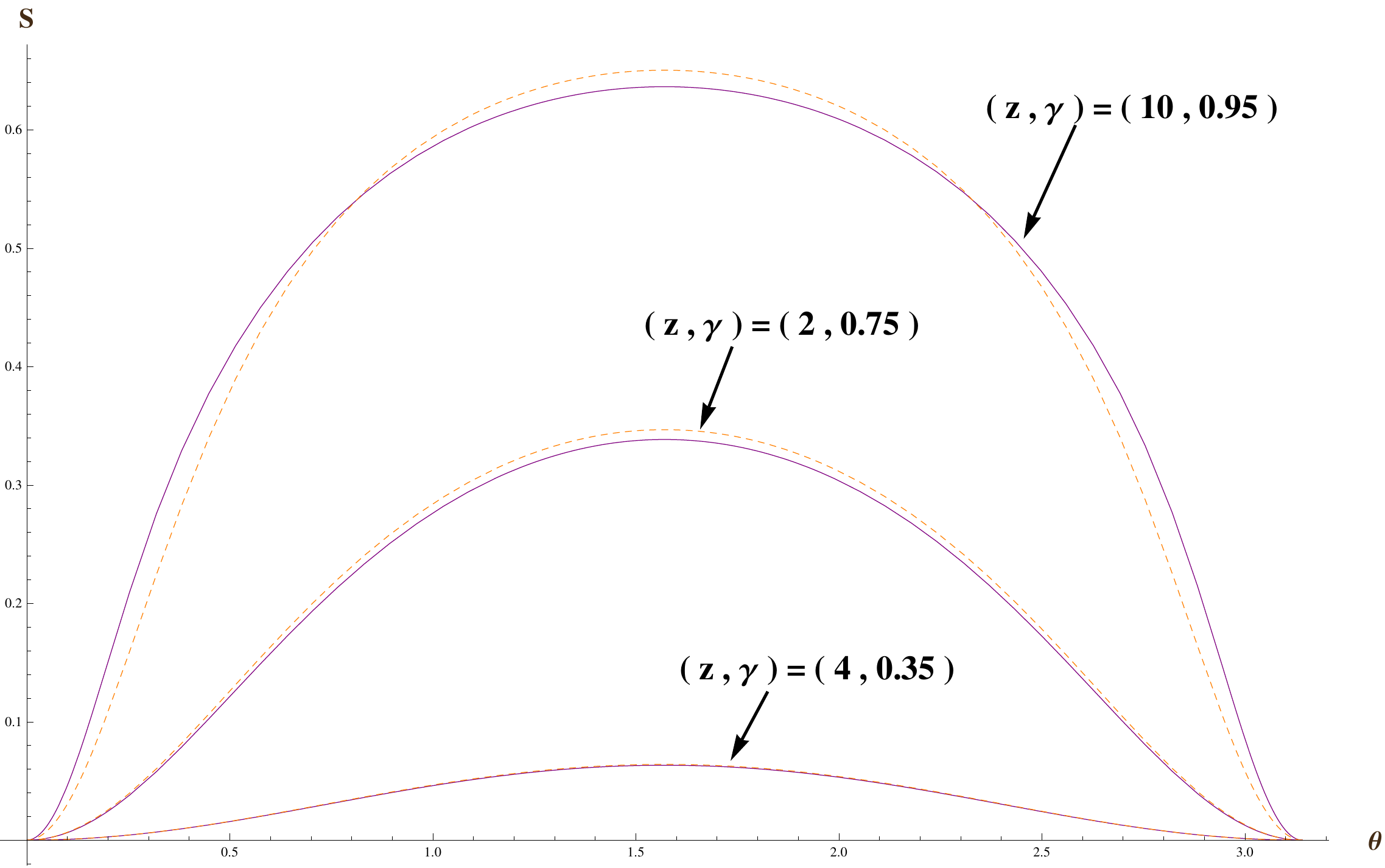}
\caption{ {\small Linear entropy as a function of the angle $\theta$ for oscillator-like (plain line) and energy-like (dashed line) SCS with different values of the amplitude $z$ and the squeezing $\gamma$.}}
\label{ChangerAngle}
\end{figure}

All our calculations have been realized with a 50:50 beam splitter, i.e., we have taken $\theta = \pi/2$ in the operator $\hat{B}(\theta)$.  In this subsection, we examined the effect on the linear entropy of changing the angle $\theta$ of the beam splitter for both types of SCS and for different values of the amplitude $z$ and the squeezing $\gamma$. Fig \ref{ChangerAngle} shows that the level of entanglement is at its maximum for the symmetric beam splitter ($\theta = \pi/2$). It is in agreement with what has been observed for the harmonic oscillator \cite{Kim1}.

\section {Conclusions \label{sec:conclusions}}

In this paper, following the idea used for the harmonic oscillator, we have measured the level of entanglement of SCS of the Morse potential using a beam splitter. We have considered two types of such states and a symmetric beam splitter. We have used the linear entropy to measure the level of entanglement since it is a good approximation of the von Newman entropy, which is mainly used in the recent literature. 

We started the paper giving the general construction of SCS associated with deformed Heisenberg algebras for the case of the harmonic oscillator and the Morse potentials. Our deformed states are different from theses considered in preceding approaches. Moreover, these approaches were mainly focused on such states but with no squeezing. 

Linear entropy has been computed first for different SCS of the harmonic oscillator giving expected results with respect to the behaviour of the usual CS, in particular. Second, the two types of SCS of the Morse potential that have been constructed in our preceding paper lead to interesting properties with respect to localization and dispersion \cite{angelova12}. We showed that squeezing was always present, even if we consider states for which the squeezing parameter is zero. With respect to the measure of entanglement, we have been able to confirm this fact. For small values of the amplitude $z$, both states create almost the same quantity of entanglement but while it stays stable in the case of energy -like SCS, it increases suddenly in the oscillator-like SCS. We have shown that these results are in agreement with the behaviour of the density probability of our states. We have also seen that Morse SCS are more non classical than those of the harmonic oscillator since they always create some entanglement. Time evolution of the linear entropy has shown periodicity with a minimum at the initial state ($t=0$) and the oscillations are different for different types of states. Finally, we have observed that the symmetric beam splitter is the one that creates a maximum of entanglement like in the case of the harmonic oscillator.


\section*{Acknowledgments}
VH acknowledges the support of research grants from NSERC of Canada.


\end{document}